\documentclass[preprint,5p]{elsarticle}

\usepackage{amsmath}%,amssymb}

\usepackage{amssymb}
\allowdisplaybreaks

\usepackage{makeidx}
\usepackage{amsfonts}
\usepackage[usenames,dvipsnames]{pstricks}
\usepackage{subfigure}
\usepackage{epsfig}
\usepackage{pst-grad} % For gradients
\usepackage{mathtools}
\usepackage[colorlinks,hyperindex]{hyperref}
\usepackage{array}
\hypersetup
{	colorlinks,%
	citecolor=black,%
	linkcolor=black,%
	urlcolor=black,%
}

\usepackage{allpurpose}

\usepackage{flushend}% Align the end of the last page.

\newcommand\nn{{\nonumber}}

\begin{document}

\title{Universal time delay in static spherically symmetric spacetimes for null and timelike signals}

\author[1]{Haotian Liu}
\ead{htliu@whu.edu.cn}

\author[1]{Junji Jia\corref{cor1}}
\ead{%Corresponding author:
junjijia@whu.edu.cn}

\cortext[cor1]{Corresponding author}
\address[1]{MOE Key Laboratory of Artificial Micro- and Nano-structures, School of Physics and Technology, Wuhan University, Wuhan, 430072, China}

%\address[2]{Center for Astrophysics \& MOE Key Laboratory of Artificial Micro- and Nano-structures, School of Physics and Technology, Wuhan University, Wuhan, 430072, China}

\date{\today}

\begin{abstract}
A perturbative method to compute the total travel time of both null and lightlike rays in arbitrary static spherically symmetric spacetimes  in the weak field limit is proposed. The resultant total time takes a quasi-series form of the impact parameter. The coefficient of this series at a certain order $n$ is shown to be determined by the asymptotic expansion of the metric functions to the order $n+1$.  To the leading order(s), the time delay, as well as the difference between the time delays of two kinds of relativistic signals, is then shown to take a universal form for all SSS spacetimes. This universal form depends on the mass $M$ and a post-Newtonian parameter $\gamma$ of the spacetime. The analytical result is numerically verified using the central black hole of M87 as the gravitational lensing center.
\end{abstract}


\maketitle

\section{Introduction}

Nowadays the time delay of gravitational lensed images by compact object, galaxies or their clusters has become a powerful tool in astrophysics and cosmology \cite{Refsdal:1964nw}.
For lensing by compact objects, the time delay can be used to constrain properties of the spacetime, such as its mass or naked singularities \cite{Virbhadra:2007kw,Eiroa:2013nra}. While time delays by galaxies or galaxy cluster can independently and accurately measure the Hubble parameter to percentage level, to constrain the lens mass profile, the line-of-sight mass distribution, dark matter substructures and the dark Universe parameters
\cite{Keeton:2008gq,Coe:2009wt,Linder:2011dr,Suyu:2013kha,Bonvin:2016crt,Treu:2016ljm,Liao:2017ioi}.

Traditionally, the time delay has always been obtained from light spectral data.
With the discovery of extragalactic neutrinos \cite{Hirata:1987hu, Bionta:1987qt,IceCube:2018dnn,IceCube:2018cha} and the gravitational waves (GWs) \cite{Abbott:2016blz,Abbott:2016nmj,Abbott:2017oio,TheLIGOScientific:2017qsa,Monitor:2017mdv}, and especially the lensed supernovas \cite{Kelly:2014mwa,Goobar:2016uuf} and simultaneous observation of GW+GRB events \cite{TheLIGOScientific:2017qsa,Monitor:2017mdv}, it is clear that both neutrinos and GWs can act as messengers for the time delay effect. Although it is known that neutrinos \cite{Tanabashi:2018oca} as well as GWs in some gravitational theories beyond GR \cite{Sakstein:2017xjx,Baker:2017hug} have nonzero masses, in previous considerations of their time delays the formula for null rays are used \cite{Fan:2016swi,Liao:2017ioi,Wei:2017emo,Yang:2018bdf}. Since the subluminal speed of these massive particles can make extra contribution to the time delay, one should compute using timelike geodesics rather than null ones if high accuracy is desired.

Previously, we have shown that the time delay in the Schwarzschild spacetime for null or timelike rays receives factors of correction \cite{Jia:2019hih}. In considering other spacetimes, however, the effect of the spacetime parameters such as electromagnetic charges, angular momentum and other effective charges such as in the Bardeen \cite{Bardeen:1968,AyonBeato:2000zs},  Janis-Newman-Winicour \cite{Janis:1968zz} and Einstein-Born-Infeld spacetimes \cite{Breton:2002td,Eiroa:2005ag} etc. on the time delay is still unclear.
In this work, we present a perturbative method to calculate the total travel time and time delay in arbitrary static spherically symmetric (SSS) spacetimes for null and timelike signals with general velocity. The result of the total time takes a quasi-series form of the impact parameter $b$, and the time delay to the leading order(s) takes a universal form, depending on leading expansion coefficients of two metric functions.
We use the geometric unit $G=c=1$ throughout the letter.

\section{Total travel time in SSS spacetimes}

The most general SSS metric can be described by
\be
\dd s^2=-A(r)\dd t^2+B(r)\dd r^2+C(r)\lb \dd \theta^2+\sin^2\theta \dd \varphi^2\rb \label{sphmetric}
\ee
where $(t,~r,~\theta,~\varphi)$ are the coordinates and $A,~B,~C$ are metric functions depending on $r$ only.
It is routine to find the geodesic equations in this metric 
\bea
&&\dot{t}=\frac{E}{A},\\
&&\dot{\phi}=\frac{L}{C}, \label{eq:phieq}\\
&&\dot{r}^2=\frac{(E^2-\kappa A)C-L^2A}{ABC},\label{eq:req}
\eea
in which $\kappa=0,~1$ for null and timelike signals respectively. Here $L$ and $E$ are first integral constants representing respectively the angular momentum and energy of the null ray or the unit mass of the timelike particle.
Note that due to the spherical symmetric, we will always set $\theta=\pi/2$ for the geodesic motion without losing any generality. 
Using Eqs. \eqref{eq:phieq} and \eqref{eq:req}, one can obtain  the equation of motion for $\dd t/\dd r$. Further integrating it from the source located at radius $r_s$ to the closest radius $r_0$ and then to a detector at $r_d$, one obtains the total travel time $t$ of the signal,
\be
t=\lsb  \int^{r_s}_{r_0}\!+ \! \int^{r_d}_{r_0} \rsb \frac{E \sqrt{B C} }{LA} \frac{LA}{\sqrt{A\lsb \lb E^2 - \kappa A\rb C - L^2 A \rsb}}\dd r.
\label{eq:ttotaldef}
\ee
Note that we kept some terms un-canceled in the numerator and denominator for the later convenience in Eqs. \eqref{eq:drtrans} and \eqref{eq:sinttrans}. 
In this letter, we will concentrate on asymptotically flat spacetimes, in which $L$ and $E$ are related to the velocity $v$ at infinity and impact parameter $b$ by
\be
|L|=|\mathbf{r}\times \mathbf{p}|=\frac{v}{\sqrt{1-v^2}}b,~ E=\frac{1}{\sqrt{1-v^2}}. \label{eq:ledef}\ee
The angular momentum $L$ can also be related to $r_0$ using the radial   equation of motion $\dd r/\dd t|_{r=r_0}=0$, to find
\be
|L| = \sqrt{C(r_0) \lsb E^2 - \kappa A(r_0) \rsb /A(r_0)}. \label{angmem}
\ee
Further using Eqs. \eqref{eq:ledef} and \eqref{angmem}, one then can establish a relation between the impact parameter $b$ and $r_0$
\bea
\frac{1}{b}&=&\frac{\sqrt{E^2-\kappa}}{\sqrt{E^2-\kappa A(r_0)}}\sqrt{\frac{A(r_0)	}{C(r_0)}}\label{eq:bequal}\\
&\equiv& p\lb \frac{1}{r_0}\rb,\label{eq:pdefs}
\eea
where in the last step we denoted $1/b$ as a function $p$ of $1/r_0$.

The key to proceed is to do a special change of variable in the total time formula \eqref{eq:ttotaldef}, after which we can do a series expansion of the impact parameter and then prove rigorously the integrability of the expansion for null and timelike rays in arbitrary SSS spacetimes.
The change of variable simply utilize the inverse function of $p(x)$, which we denote as $q(x)$, so that $r$ is changed to $u$ through relation
\be
\frac{1}{r}=q\lb \frac{u}{b}\rb, \label{eq:udef}\ee
or equivalently 
\bea 
u&=&b\cdot p\lb \frac{1}{r}\rb\nn\\
&=&\sqrt{\frac{A(r)C(r_0)\lsb E^2-\kappa A(r_0)\rsb }{A(r_0)C(r)\lsb E^2-\kappa A(r)\rsb }} \label{eq:uequals}
\eea
where in the second step Eqs. \eqref{eq:bequal} and \eqref{eq:pdefs} are used.
Now using Eq. \eqref{eq:udef} together with Eq. \eqref{eq:ledef}, the infinitesimal and the first fractional term in the integrand of Eq. \eqref{eq:ttotaldef} become respectively
\bea
&&\dd r \to -\frac{1}{p^\prime ( q) q^2}\frac{1}{b}\dd u. \label{eq:termstrans}\\
&&\frac{E \sqrt{B(r) C(r)} }{LA(r)} \to \frac{\sqrt{B(1/q)C(1/q)}}{A(1/q)} \frac{1}{bv},\label{eq:drtrans}
\eea
Using Eqs. \eqref{angmem} and \eqref{eq:uequals}, the second fractional term in the integrand of Eq. \eqref{eq:ttotaldef} becomes
\be
\frac{LA(r)}{\sqrt{A(r)\lsb \lb E^2 -\kappa A(r)\rb C(r) - L^2 A(r)\rsb } }
\to \frac{u}{\sqrt{1-u^2}}. \label{eq:sinttrans}
\ee
Note on the right hand sides of Eqs. \eqref{eq:drtrans} to \eqref{eq:sinttrans} , $q$ is $q(u/b)$ and $p^\prime$ is the derivative of function $p$.
The integral limits of Eq. \eqref{eq:ttotaldef} should change to the limits of variable $u$, i.e.,
\be
r_0\to 1,~r_{s,d}\to b\cdot p\lb \frac{1}{r_{s,d}}\rb.
\ee
Collecting these together, the total time \eqref{eq:ttotaldef} becomes
\be
t=\lsb \int_{b\cdot p\lb \frac{1}{r_s}\rb}^1+\int_{b\cdot p\lb \frac{1}{r_d}\rb }^1\rsb y\lb \frac{u}{b}\rb  \frac{\dd u}{u\sqrt{1-u^2}} \label{eq:idef3}
\ee
where
\be y\lb \frac{u}{b}\rb =
\frac{\sqrt{B(1/q)C(1/q)}}{A(1/q)} \frac{u}{b}\frac{1}{v}
\frac{1}{p^\prime ( q) q^2}
\frac{u}{b} .\label{eq:ydef}
\ee

It is then essential to note that this function $\displaystyle  y\lb \frac{u}{b}\rb$ depends on $u$ only through the ratio $\frac{u}{b}$. Using the definition \eqref{eq:ydef}, when the metric functions $A(r),~B(r)$ and $C(r)$ are known, then $\displaystyle  y\lb \frac{u}{b}\rb$ can be series expended in the powers of $u/b$ to yield the result
\be
y\lb \frac{u}{b}\rb=\sum_{n=-1}^{\infty} y_n \lb \frac{u}{b}\rb^n, \label{eq:yexp}
\ee
where $y_n$ are the expansion coefficients. Note the explicit form of the inverse function $q(x)$ is not actually needed in this process since one can use the Lagrange inverse theorem and function $p(x)$ to find its expansion.
To obtain the general form of $y_n$, we will use the following asymptotic expansion of the metric functions
\be
A(r) =1+\sum_{n=1} \frac{a_n}{r^n},\ B(r) =1+\sum_{n=1} \frac{b_n}{r^n},\ \frac{C(r)}{r^2}=1+\sum_{n=1} \frac{c_n}{r^n}, \label{eq:abcform}
\ee
where $a_n$, $b_n$ and $c_n$ are finite constants. Without losing any generality, the constant $a_1$ will latter be identified with the ADM mass $M$ of the spacetime $(a_1=-2M)$, and the constant $b_1$ is conventionally referred as the $\gamma$ parameter $(b_1=2M\gamma)$ in the parameterized post-Newtonian (PPN) formalism of gravity  \cite{Weinberg:1972kfs}.
Substituting these into Eq. \eqref{eq:ydef} and carrying out the power expansion of $u/b$, the first three orders of $y_n$ are found to be
\begin{align}
&y_{-1} = \frac{1}{v}, \nonumber \\
&y_0 = \frac{1}{2v}\lb \frac{a_1}{v^2}-2a_1+b_1\rb, \label{eq:asyyn} \\
&y_1 = \frac{8 a_1^2-4 (b_1+c_1) a_1-\lb b_1-c_1\rb^2-8 a_2+4 b_2+4 c_2}{8v}. \nn
\end{align}
It is seen that for the order $n$ coefficient $y_n$, only the coefficients up to order $n+1$ of the metric functions contribute.

Substituting Eq. \eqref{eq:yexp} into the total time \eqref{eq:idef3} and do a further change of variable $u=\sin\theta$ which is suggested by the denominator $\sqrt{1-u^2}$,
this becomes
\be
t=\sum_{n=-1}^\infty \frac{y_n}{b^n} \lsb \int_{\beta_s}^{\frac{\pi}{2}} +\int_{\beta_d}^{\frac{\pi}{2}} \rsb \sin^{n-1} \theta\dd\theta
\label{eq:idef35}
\ee
where we denoted respectively\be
\beta_{s,d} \equiv\arcsin(b\cdot p(1/r_{s,d})).
\label{eq:betadef}\ee
At this point, the integrability of Eq. \eqref{eq:idef35} to any desired order of $y_n/b^n$ becomes clear because the integration part can always be carried out to yield
\begin{align}
l_n(\beta_s,~\beta_d)&\equiv \lsb \int_{\beta_s}^{\frac{\pi}{2}} +\int_{\beta_d}^{\frac{\pi}{2}} \rsb\sin^{n-1} \theta\dd \theta \nn \\
&=\sum_{i=s,d}
\begin{cases}
\displaystyle \cot\beta_i,&~n=-1\\
\displaystyle \ln\lsb\cot\lb\frac{\beta_i}{2}\rb\rsb,&~n=0
\end{cases} ,\label{eq:cndef}
\end{align}
and for odd and even positive $n$ respectively \cite{bk:inttable}
\begin{align}
&l_n(\beta_s,~\beta_d)=\sum_{i=s,d} \frac{(n-2)!!}{(n-1)!!}\times \tag{\ref{eq:cndef}} \\
&\begin{cases}
\displaystyle  \left(\frac{\pi}{2}-\beta_i
    +\cos\beta_i\sum_{j=1}^{[\frac{n-1}{2}]} \frac{(2j-2)!!} {(2j-1)!!}\sin^{2j-1} \beta_i\right),\\
    \hspace{5.8cm}n=2k+1,\\
\displaystyle  \cos\beta_i \left(1
    +\sum_{j=1}^{[\frac{n-1}{2}]} \frac{(2j-1)!!}{(2j)!!} \sin^{2j}\beta_i\right),~n=2k,~k\in\mathbb{N}.
\end{cases} \nn
\end{align}
Therefore, the total time \eqref{eq:idef35} becomes
\be
t=\sum_{n=-1}^\infty \frac{y_n}{b^n} l_n(\beta_s,\beta_d) \label{eq:idef4}
\ee
where $y_n$ are given in Eq. \eqref{eq:asyyn} and $l_n$ in Eq. \eqref{eq:cndef}.
This is the total travel time applicable to all SSS spacetimes and both null and timelike rays, and moreover for both large and smaller impact parameters $b$.

For practical gravitational lensing (GL) observation, the relation $r_{s,d}\gg b$ is satisfied. Moreover, we will assume that the expansion parameters of the metric functions in Eq. \eqref{eq:abcform} satisfies the weak field limit, i.e.,  $\mathcal{O}(b^n)\gg |a_n|,~|b_n|$ or $|c_n|$. In these two limits, we can expand the total time \eqref{eq:idef4} in the powers of $\frac{1}{b}$ and $\frac{b}{r_{s,d}}$. To this end, we can expand the first three $l_n$ $(n=-1,0,1)$ in Eq. \eqref{eq:cndef} to the order of $\lb \frac{M}{b}\rb^1$ and $\lb \frac{b}{r_{s,d}}\rb^1$
\begin{align}
&l_{-1} =\sum_{i=s,d} \lsb \frac{r_i}{b}-\frac{b}{2 r_i} +\lb \frac{c_1}{2}-\frac{a_1}{2 v^2}\rb \frac{1}{b} + \mathcal{O}\lb \frac{b^2}{r_i^2}, \frac{M^2}{b^2}\rb \rsb, \nonumber \\
&l_0 = \sum_{i=s,d} \lsb -\ln \lb\frac{b}{2 r_i}\rb + \frac{c_1 v^2-a_1}{2 v^2}\frac{b}{r_i} \frac{1}{b} + \mathcal{O}\lb \frac{b^2}{r_i^2}, \frac{M^2}{b^2}\rb \rsb, \nonumber \\
&l_1 =\sum_{i=s,d}\lsb \frac{\pi}{2}-  \frac{b}{r_i}+ \mathcal{O}\lb \frac{b^2}{r_i^2}, \frac{M^2}{b^2}\rb \rsb . \label{eq:asyln}
\end{align}
Substituting this expansion and Eq. \eqref{eq:asyyn} into Eq. \eqref{eq:idef4}, the total time becomes
\begin{align}
t =&\sum_{i=s,d}\lcb \frac{r_i}{v}-\frac{b^2}{2 r_iv}+\lb \frac{a_1}{2 v^3}-\frac{2a_1-b_1}{2v}\rb
 \ln \frac{2 r_i}{b} \right.\nn\\
 &+\lb \frac{c_1}{2}-\frac{a_1}{2 v^2}\rb \frac{1}{v} +\frac{\pi}{16bv}\times \nonumber \\
&\lsb 8 a_1^2-4 (b_1+c_1) a_1-\lb b_1-c_1\rb^2-8 a_2+4 b_2+4 c_2\rsb \nonumber \\
& +\frac{1}{r_i}\lsb -\frac{a_1^2}{4v^5}+\frac{a_1}{4v^3}(2a_1-b_1+c_1)\right. \nn\\
&\left.+\frac{1}{8v}\lb 4a_1(-2a_1+b_1)+b_1^2+c_1^2+8a_2-4b_2-4c_2\rb \rsb  \nn\\
&\left.+\mathcal{O}\lb \frac{b^3}{r_{s,d}^2},\frac{M^2}{b}\rb\rcb \label{eq:ttgexp}
\end{align}
where all terms are arranged in an decrease order.
The first, second and fourth terms, i.e., the $\mathcal{O}(r_i)$, $\mathcal{O}(b^2/r_i)$, $\mathcal{O}(M)$ order terms, are from $l_{-1}$ in Eqs. \eqref{eq:asyln}. Similarly, the third and part of the sixth terms, which are of order $\mathcal{O}(\ln(r_i/b))$ and $\mathcal{O}(M^2/r_i)$ respectively, originate from $l_0$. While the fifth and the rest of the sixth order terms, of order $\mathcal{O}(M^2/b)$ and $\mathcal{O}(M^2/r_i)$, are from $l_1$. As we will show next, expansion to these orders are more than enough to find leading order(s) useful for observations.

\section{Time delay in GL}

\begin{figure}[htp]
\includegraphics[width=0.45\textwidth]{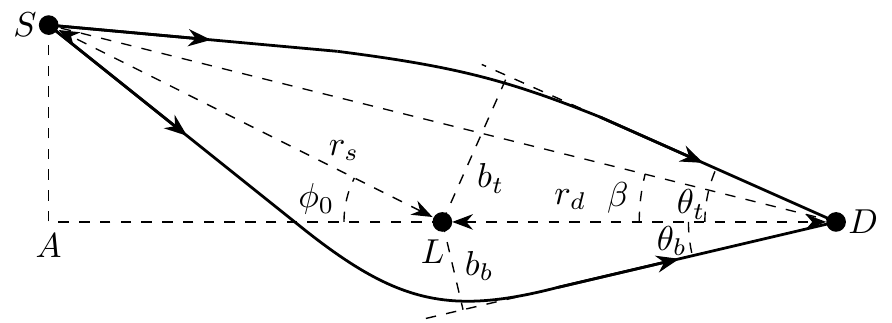}
\caption{The GL in an SSS spacetime. $S,~L,~D$ are the source, lens and detectors. $b_b$ and $b_t$ are the impact parameters for the bottom and top paths respectively. \label{fig:glconf}}
\end{figure}

Using the total time \eqref{eq:ttgexp}, we can compute the time delay between images of the same source in GL. To do this, we first need to find the total time of each image separately. Let us suppose that the source, lens and detector are in a configuration described in Fig. \ref{fig:glconf}, where $\phi_0$ is the angle of the lens-source direction against the lens-detector axis. $\beta$ denotes the angle of the detector-source direction against the lens-detector axis in the no lens limit. Using  triangles $\triangle ASL$ and $\triangle ASD$, apparently we can establish the geometric relation
\be
\lb r_d+r_s\cos\phi_0\rb \sin\beta=r_s\sin\phi_0 \label{eq:phi0betarel}\ee
between the two angles $\phi_0$ and $\beta$ and the source, detector radii $r_s$ and $r_d$ respectively.

Clearly, for the two paths of any given two images, the only parameter that are different in the total time formula \eqref{eq:ttgexp} or \eqref{eq:idef4} is their impact parameter $b$. Therefore we need to find a way to compare these two $b$'s in order to compute their time delay. Here we will avoid as much as possible any non-exact equations or equations whose truncation errors are hard to track, such as the approximate equation $b\approx r_d \theta$ where $\theta$ is the apparent angle of the images. Rather, in this letter we will use a more exact and trackable method to link the two impact parameters.

From Ref. \cite{ourletter}, the change of the angular coordinates $\Delta \varphi$ from the source to the detector in an SSS spacetime has been computed as a series form to high order of $M/b$. For the purpose of time delay, we will only use this series to the first order of $M/b$ and $b/r_{s,d}$ (henceforth we use $M$ and $\gamma$ to completely replace $a_1$ and $b_1$), i.e.,
\be
\Delta \varphi(b)\approx \pi + \frac{2M}{b}\lb \gamma+\frac{1}{v^2}\rb -b \lb \frac{1}{r_s}+\frac{1}{r_d}\rb .
\label{eq:alphalbsss}
\ee
We emphasis that to this order, among all expansion coefficients in Eq. \eqref{eq:abcform}, $\Delta \varphi(b)$ only depends on mass $M$ and $\gamma$ but not high order $a_n,~b_n~(n=2,\cdots)$ or any of $c_n~(n=1,2,\cdots)$. One will see later that this point contributes to the universality of the time delay to the leading order(s) in Eq. \eqref{eq:detbfbeta}.

For GL shown in Fig. \ref{fig:glconf}, the change of the angular coordinates from the bottom and top sides are respectively $\pm \pi + \phi_0$. Equating them to Eq. \eqref{eq:alphalbsss} with $b$ substituted by impact parameters $b_{t,b}$ from both sides, we have
\be
\mp\pi + \phi_0=\mp \Delta \varphi(b_{t,b}) .
\ee
where the sign on the top (or bottom) is for the top (or bottom) path respectively. From this, we can solve $b_{t,b}$ as
\begin{align}
b_{t,b}=&\frac{\phi_0 r_d r_s }{2 (r_d+r_s)} \lsb \sqrt{1+\eta}\pm 1\rsb
\label{eq:blrtophi}
\end{align}
where
\be \eta= \frac{8M\left(1 +\gamma v^2\right) (r_d+r_s)}{\phi_0^2v^2 r_d r_s} .\ee Eq. \eqref{eq:blrtophi} establishes a simple and yet accurate relation between the angle $\phi_0$ and the impact parameters, with the only approximation tracking back to the truncation of the change of the angular coordinate \eqref{eq:alphalbsss}.

Substituting these two impact parameters \eqref{eq:blrtophi} and $a_1=-2M$ into the total time \eqref{eq:ttgexp} and subtracting each other, we see that the leading, fourth and sixth terms of Eq. \eqref{eq:ttgexp}, i.e., the $\mathcal{O}(r_i),~\mathcal{O}(M)$ and $\mathcal{O}(M^2/r_i)$ order terms, are exactly canceled between the two total times. The second, third and fifth terms, emerged as the respect leading term in the expansion of $l_{-1},~l_0$ and $l_1$ in Eq. \eqref{eq:asyln}, becomes respectively the three terms in the following result for the time delay
%\begin{widetext}
\begin{align}
&\Delta t
=\frac{4M\left(1 +\gamma v^2\right)}{v^3 \eta}\sqrt{1+\eta}\nn\\
& +\frac{2M \lsb 1- v^2(2 +\gamma)\rsb}{v^3} \ln \lb 1-\frac{2}{1+\sqrt{1+\eta}}\rb +\pi  \phi_0 v\times\nonumber \\
&\frac{ \lsb -2 a_2+b_2+M^2(8+4\gamma-\gamma^2)\rsb} {4M\left(1+\gamma v^2\right)} + \mathcal{O}\lb \frac{b^3}{r_{s,d}^2},\frac{M^2}{b}\rb  . \label{eq:detbf}
\end{align}
%\end{widetext}
It is then easy to see that when
\be
\eta\gg 1, ~\mbox{i.e.,}~\frac{M}{r_{d,s}}\gg \phi_0^2, \ee
the logarithmic term of Eq. \eqref{eq:detbf} will take a form of $\ln(1+\text{ a small quantity})$ and a Taylor expansion could show that this term is comparable to the first term of Eq. \eqref{eq:detbf}. Moreover, it can also be shown that in this case the first two terms will be much larger than the third one, which therefore can be ignored. The time delay to the leading order in this case then takes a simple form
\begin{align}
\Delta t
=\frac{8M(1+\gamma)}{v\sqrt{\eta}}+\mathcal{O}\lsb M\lb \frac{r_{d,s}}{M}\phi_0^2\rb^{3/2},\phi_0M, \frac{b^3}{r_{s,d}^2},\frac{M^2}{b}\rsb.
\label{eq:tdlargeeta}
\end{align}
On the other hand, if $M/r_{d,s}\ll \phi_0^2$, then the first term will dominate the second which in turn is much larger than the third one. In this limit, the time delay can also be expanded to yield a simple result to the leading order
\be
\Delta t = \frac{4M(1+\gamma v^2)}{ v^3 \eta}+ \mathcal{O}\lb M,\frac{b^3}{r_{s,d}^2}\rb. \label{eq:tdsmalleta}
\ee
When $\gamma =1$, as for many SSS spacetimes including the most common Schwarzschild and Reisnerr-Nordstrom etc., the time delay \eqref{eq:tdlargeeta} and \eqref{eq:tdsmalleta} reduce to Eq. (35) and (32) respectively of Ref. \cite{Jia:2019hih}.

Combining the above two limits, it is clear then in any case, the time delay can always be well described by the sum of first two terms of Eq. \eqref{eq:detbf}.
In GL computations, the source angular position is often represented by $\beta$ in Fig. \ref{fig:glconf} rather than $\phi_0$. To this end, one can simply solve $\phi_0$ in terms of $\beta$ from Eq. \eqref{eq:phi0betarel} and substitute in to the first two terms of Eq. \eqref{eq:detbf}, yielding
\begin{align}
\Delta t
=&\frac{4M\left(1 +\gamma v^2\right)}{v^3 \eta(\beta,v)}\sqrt{1+\eta(\beta,v)}\nn\\
& +\frac{2M\lsb 1 - v^2(2 +\gamma)\rsb}{v^3} \ln \lb 1-\frac{2}{1+\sqrt{1+\eta(\beta,v)}}\rb \nonumber \\
& + \mathcal{O}\lb \beta M, \frac{b^3}{r_{s,d}^2},\frac{M^2}{b}\rb \label{eq:detbfbeta}
\end{align}
where $\eta(\beta,v)$ is
\be
\eta(\beta,v)=\frac{8M\left(1+\gamma v^2\right)r_s}{\beta^2v^2(r_d+r_s)r_d}. \label{eq:etabetadef}
\ee

A few features of this result is remarkable. First, only two parameters from an SSS spacetime, the mass $M$ and the PPN parameter $\gamma$ of the metric function $B(r)$, appear in the time delay to these leading order(s). All other quantities in Eq. \eqref{eq:detbfbeta}, i.e., $r_{s,d},~v$ and $\beta$, are geometric or kinetic variables associated with the initial/final conditions of the signal particle. High order spacetime parameters $a_n,~b_n~(n=2,3,\cdots)$ and all of $c_n~(n=1,2,\cdots)$ in Eq. \eqref{eq:abcform}, including (effective) charges etc., have little effect on the time delay of weak-deflection GL. Because of this, all SSS spacetime time delays at the leading order(s) take a universal form described by Eq. \eqref{eq:detbfbeta}. Secondly, the first term of Eq. \eqref{eq:detbfbeta} originates from the $l_{-1}$ term of the total time \eqref{eq:idef4}, which represents the geometric propagation time. Meanwhile, the second term of Eq. \eqref{eq:detbfbeta} can be traced back to the $l_0$ term, which corresponds to half of the conventional Shapiro time delay \cite{Jia:2019hih}. Therefore the analysis in this section shows that in any circumstance of the lensing parameters, i.e. small or large $\eta(\beta,v)$, to find the weak field time delay one only needs to calculate to the 2nd non-trivial order of the total time \eqref{eq:idef4}. In other words, only $y_{-1,0}$ and $l_{-1,0}$ are needed.

\section{Application of the results}

\begin{figure}[htp!]
\includegraphics[width=0.45\textwidth]{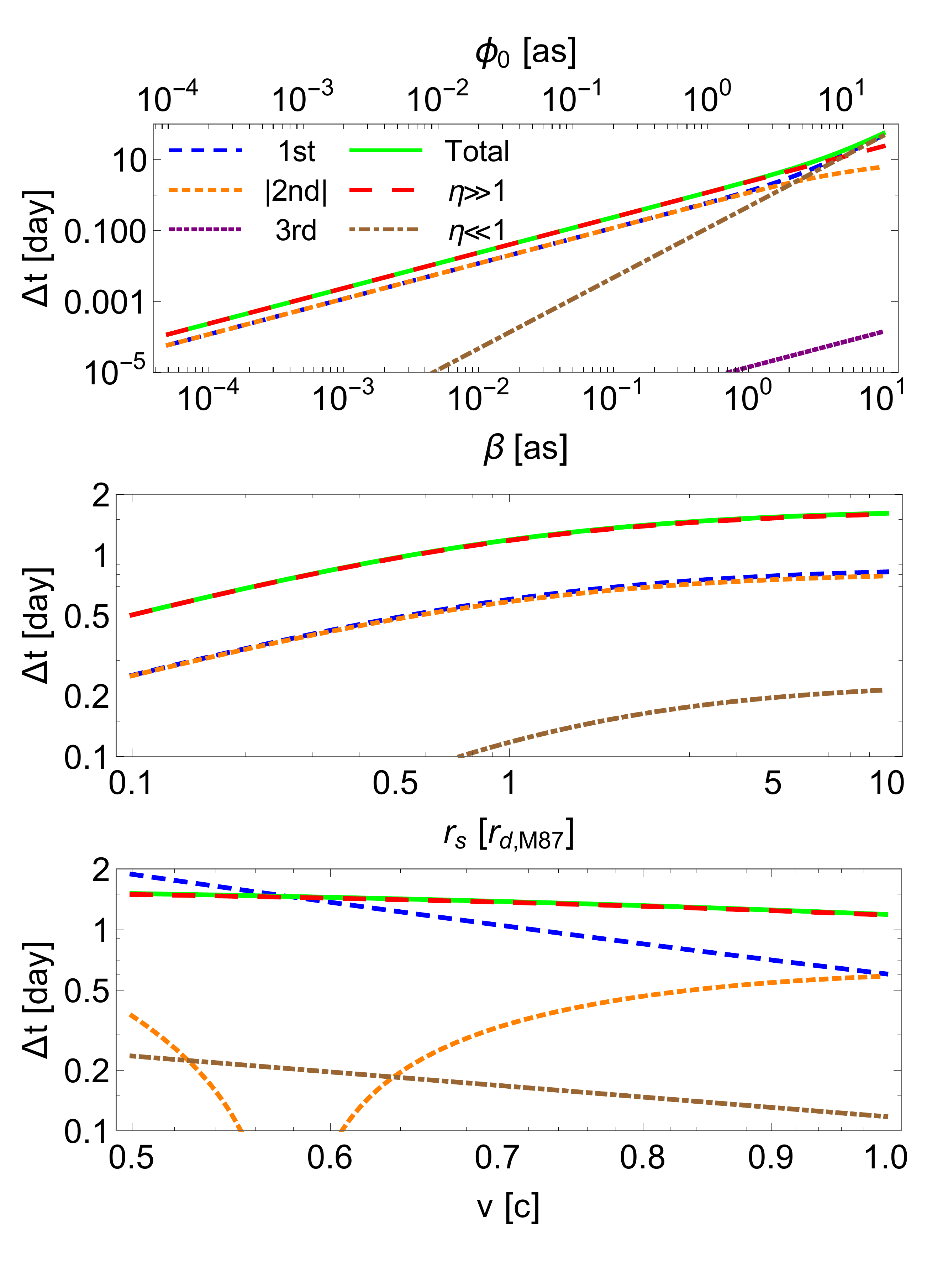}
\caption{\label{fig:tdm87one} The time delay caused by M87 supermassive BH, as a function of $\beta$ (or $\phi_0$) (top), $r_s$ (middle) and $v$ (bottom). The parameters chosen are $r_s=r_d$, $\phi_0=1$ [as] and $v=c$, except the parameter that is varied in the $x$-axis of each subplot. The 1st, |2nd|, 3rd and ``total'' curves correspond to the first, absolute value of the second, third and all three terms of Eq. \eqref{eq:detbf} respectively. The $\eta\gg 1$ and $\eta\ll 1$ curves corresponds to Eqs. \eqref{eq:tdlargeeta} and \eqref{eq:tdsmalleta}. }
\end{figure}

To check the validity of the time delay Eqs. \eqref{eq:detbf}-\eqref{eq:etabetadef}, we apply these results to the supermassive BH in the center of galaxy M87, which we model as an SSS with $\gamma=1$ but otherwise arbitrary spacetime. Note that here we do not need to specify the exact type of the spacetime because as shown in Eq.  \eqref{eq:detbfbeta}, the time delay depends to the leading order only on $M$ and $\gamma$. Using the mass $M=6.5\times 10^6M_\odot$ and $r_d=16.8$ [Mpc] \cite{Akiyama:2019cqa} for M87, we plotted the time delay \eqref{eq:detbf} in Fig. \ref{fig:tdm87one} as a function of other parameters.

We see from the first subplot that as $\beta$ increases, the first and second terms of Eq. \eqref{eq:detbf} (or \eqref{eq:detbfbeta}) are of similar values until about $\beta\approx  2$ [as], beyond which the first term dominates the second one. Moreover, the first two terms are much larger than the third one in the whole range. When $\beta \lesssim 4$ [as] and $\beta \gtrsim 4$ [as] respectively, the total time delay approaches the $\eta\gg 1$ and $\eta \ll 1$ limits respectively.
In the second subplot, in the whole range of $r_s$, the first and second terms of Eq. \eqref{eq:detbf} (or \eqref{eq:detbfbeta}) are comparable, and their combination becomes the total time delay which is approximated by the $\eta\gg 1$ limit. While in the last subplot as $v$ varies from $0.5c$ to $c$, the value of the first term decreases to that of the second term, which increases from negative to positive. Again, their combination forms the total time delay which is approximated by the $\eta\gg 1$ limit. One can verify that for these parameters or parameter ranges, all the above described behaviors match the prediction of Eqs. \eqref{eq:detbf}-\eqref{eq:etabetadef} perfectly.

Since both the supernovas and GW+GRB events emit two kinds of relativistic signals (almost) simultaneously, people have proposed to use the difference between the time delays of both kind of signal to constrain properties of the signals \cite{Fan:2016swi,Jia:2019hih}. Using Eq. \eqref{eq:detbfbeta} for signals with velocity $v_1$ and $v_2$ respectively, the time delay difference becomes
\begin{align}
\Delta^2 t
=&\lsb \frac{4M(1+\gamma )\sqrt{1+\eta(\beta,1)}}{\eta(\beta,1)}\right.\nn\\
&\left.+2M(1-\gamma)\ln \lb 1-\frac{2}{1+\sqrt{1+\eta(\beta,1)}}\rb \rsb\Delta v \nonumber \\
& + \mathcal{O}\lb \beta \sqrt{Mr_{d,s}}, \beta^2r_{d,s}\rb(\Delta v)^2 \label{eq:diffdtbeta}
\end{align}
where $\Delta v=v_1-v_2$. Similar to Eq. \eqref{eq:detbfbeta}, when $\eta\ll 1$ both terms in the square bracket of Eq. \eqref{eq:diffdtbeta} are at the same order. Otherwise, the first term dominates. For spacetimes with $\gamma=1$, the second term vanishes and this reduces to Eq. (37) of Ref. \cite{Jia:2019hih}.
Result \eqref{eq:diffdtbeta} suggests that for all SSS spacetimes, the time delay difference also relies on only two parameters $M$ and $\gamma$. In all spacetimes where $\gamma=1$, then this time delay difference becomes completely equivalent to that of the Schwarzschild spacetime with the same mass. The corresponding analysis for neutrinos and GW/GRB time delay was carried out in Ref. \cite{Jia:2019hih} and therefore will not be repeated here.

\section{Conclusions}

We developed a perturbative method to compute the total travel time in any SSS spacetime for signal with general velocity. The result, Eq. \eqref{eq:idef4}, takes a quasi-series form of $1/b$. Only the first two orders of this result contribute to the leading order(s) of the time delay $\Delta t$, given in Eq. \eqref{eq:detbfbeta}, between different images of GL.  $\Delta t$ depends on the mass $M$ and PPN parameter $\gamma$ of the metric functions. This result reveals that in the weak field limit, high order parameters in asymptotic expansions of the metric functions, such as effective charges, have a much smaller effect on $\Delta t$ than $M$ and $\gamma$.
The difference of the time delays for different kinds of signals is also shown to take a universal form to the leading order(s), still determined by $M$ and $\gamma$.

It would be interesting to see whether the current method can be generalized to other kinds of spacetimes so that parameters other than $M$ and $\gamma$ can have a sizeable effect on the time delays and their difference. Such spacetimes include at least stationary axisymmetric spacetimes, asymptotically non-flat spacetimes and non-static/stationary spacetimes. We are currently working along these directions.

\medskip
We thank Dr. Nan Yang and Mr. Ke Huang for valuable discussions. This work is supported by the NNSF China 11504276 and MOST China 2014GB109004.


\begin{thebibliography}{}
% Let us set the no of total reference to 45 at most.
\bibitem{Refsdal:1964nw} S.~Refsdal,
  %``On the possibility of determining Hubble's parameter and the masses of galaxies from the gravitational lens effect,''
  Mon.\ Not.\ Roy.\ Astron.\ Soc.\  {\bf 128}, 307 (1964).
  %%CITATION = MNRAA,128,307;%%
  %473 citations counted in INSPIRE as of 21 Feb 2020

\bibitem{Virbhadra:2007kw} K.~S.~Virbhadra and C.~R.~Keeton,
  %``Time delay and magnification centroid due to gravitational lensing by black holes and naked singularities,''
  Phys.\ Rev.\ D {\bf 77}, 124014 (2008)
  doi:10.1103/PhysRevD.77.124014
  [arXiv:0710.2333 [gr-qc]].
  %%CITATION = doi:10.1103/PhysRevD.77.124014;%%
  %242 citations counted in INSPIRE as of 22 Feb 2020

\bibitem{Eiroa:2013nra} E.~F.~Eiroa and C.~M.~Sendra,
  %``Regular phantom black hole gravitational lensing,''
  Phys.\ Rev.\ D {\bf 88}, no. 10, 103007 (2013)
  doi:10.1103/PhysRevD.88.103007
  [arXiv:1308.5959 [gr-qc]].
  %%CITATION = doi:10.1103/PhysRevD.88.103007;%%
  %43 citations counted in INSPIRE as of 22 Feb 2020

\bibitem{Keeton:2008gq} C.~R.~Keeton and L.~A.~Moustakas,
  %``A New Channel for Detecting Dark Matter Substructure in Galaxies: Gravitational Lens Time Delays,''
  Astrophys.\ J.\  {\bf 699}, 1720 (2009)
  doi:10.1088/0004-637X/699/2/1720
  [arXiv:0805.0309 [astro-ph]].
  %%CITATION = doi:10.1088/0004-637X/699/2/1720;%%
  %102 citations counted in INSPIRE as of 22 Feb 2020

\bibitem{Coe:2009wt} D.~Coe and L.~Moustakas,
  %``Cosmological Constraints from Gravitational Lens Time Delays,''
  Astrophys.\ J.\  {\bf 706}, 45 (2009)
  doi:10.1088/0004-637X/706/1/45
  [arXiv:0906.4108 [astro-ph.CO]].
  %%CITATION = doi:10.1088/0004-637X/706/1/45;%%
  %50 citations counted in INSPIRE as of 22 Feb 2020

\bibitem{Linder:2011dr} E.~V.~Linder,
  %``Lensing Time Delays and Cosmological Complementarity,''
  Phys.\ Rev.\ D {\bf 84}, 123529 (2011)
  doi:10.1103/PhysRevD.84.123529
  [arXiv:1109.2592 [astro-ph.CO]].
  %%CITATION = doi:10.1103/PhysRevD.84.123529;%%
  %64 citations counted in INSPIRE as of 22 Feb 2020

\bibitem{Suyu:2013kha} S.~H.~Suyu {\it et al.},
  %``Cosmology from gravitational lens time delays and Planck data,''
  Astrophys.\ J.\  {\bf 788}, L35 (2014)
  doi:10.1088/2041-8205/788/2/L35
  [arXiv:1306.4732 [astro-ph.CO]].
  %%CITATION = doi:10.1088/2041-8205/788/2/L35;%%
  %98 citations counted in INSPIRE as of 21 Feb 2020

\bibitem{Bonvin:2016crt} V.~Bonvin {\it et al.},
  %``H0LiCOW – V. New COSMOGRAIL time delays of HE 0435−1223: $H_0$ to 3.8 per cent precision from strong lensing in a flat ΛCDM model,''
  Mon.\ Not.\ Roy.\ Astron.\ Soc.\  {\bf 465}, no. 4, 4914 (2017)
  doi:10.1093/mnras/stw3006
  [arXiv:1607.01790 [astro-ph.CO]].
  %%CITATION = doi:10.1093/mnras/stw3006;%%
  %224 citations counted in INSPIRE as of 21 Feb 2020

\bibitem{Treu:2016ljm} T.~Treu and P.~J.~Marshall,
  %``Time Delay Cosmography,''
  Astron.\ Astrophys.\ Rev.\  {\bf 24}, no. 1, 11 (2016)
  doi:10.1007/s00159-016-0096-8
  [arXiv:1605.05333 [astro-ph.CO]].
  %%CITATION = doi:10.1007/s00159-016-0096-8;%%
  %50 citations counted in INSPIRE as of 21 Feb 2020

\bibitem{Liao:2017ioi} K.~Liao, X.~L.~Fan, X.~H.~Ding, M.~Biesiada and Z.~H.~Zhu,
  %``Precision cosmology from future lensed gravitational wave and electromagnetic signals,''
  Nature Commun.\  {\bf 8}, no. 1, 1148 (2017)
  Erratum: [Nature Commun.\  {\bf 8}, no. 1, 2136 (2017)]
  doi:10.1038/s41467-017-01152-9, 10.1038/s41467-017-02135-6
  [arXiv:1703.04151 [astro-ph.CO]].
  %%CITATION = doi:10.1038/s41467-017-01152-9, 10.1038/s41467-017-02135-6;%%
  %40 citations counted in INSPIRE as of 22 Feb 2020

\bibitem{Hirata:1987hu} K.~Hirata {\it et al.} [Kamiokande-II Collaboration],
%``Observation of a Neutrino Burst from the Supernova SN 1987a,''
Phys.\ Rev.\ Lett.\ {\bf 58}, 1490 (1987).
%doi:10.1103/PhysRevLett.58.1490
%%CITATION = doi:10.1103/PhysRevLett.58.1490;%%
%1463 citations counted in INSPIRE as of 09 May 2017

\bibitem{Bionta:1987qt} R.~M.~Bionta {\it et al.},
%``Observation of a Neutrino Burst in Coincidence with Supernova SN 1987a in the Large Magellanic Cloud,''
Phys.\ Rev.\ Lett.\ {\bf 58}, 1494 (1987).
%doi:10.1103/PhysRevLett.58.1494
%%CITATION = doi:10.1103/PhysRevLett.58.1494;%%
%1268 citations counted in INSPIRE as of 09 May 2017

\bibitem{IceCube:2018dnn} M.~G.~Aartsen {\it et al.} [IceCube and Fermi-LAT and MAGIC and AGILE and ASAS-SN and HAWC and H.E.S.S. and INTEGRAL and Kanata and Kiso and Kapteyn and Liverpool Telescope and Subaru and Swift NuSTAR and VERITAS and VLA/17B-403 Collaborations],
  %``Multimessenger observations of a flaring blazar coincident with high-energy neutrino IceCube-170922A,''
  Science {\bf 361}, no. 6398, eaat1378 (2018)
  %doi:10.1126/science.aat1378
  %[arxiv:1807.08816 [astro-ph.HE]].
  %%CITATION = doi:10.1126/science.aat1378;%%
  %48 citations counted in INSPIRE as of 17 Jan 2019

\bibitem{IceCube:2018cha} M.~G.~Aartsen {\it et al.} [IceCube Collaboration],
  %``Neutrino emission from the direction of the blazar TXS 0506+056 prior to the IceCube-170922A alert,''
  Science {\bf 361}, no. 6398, 147 (2018)
  %doi:10.1126/science.aat2890
  %[arxiv:1807.08794 [astro-ph.HE]].
  %%CITATION = doi:10.1126/science.aat2890;%%
  %73 citations counted in INSPIRE as of 14 Feb 2019

\bibitem{Abbott:2016blz} B.~P.~Abbott {\it et al.} [LIGO Scientific and Virgo Collaborations],
  %``Observation of Gravitational Waves from a Binary Black Hole Merger,''
  Phys.\ Rev.\ Lett.\  {\bf 116}, no. 6, 061102 (2016)
  doi:10.1103/PhysRevLett.116.061102
  [arXiv:1602.03837 [gr-qc]].
  %%CITATION = doi:10.1103/PhysRevLett.116.061102;%%
  %4782 citations counted in INSPIRE as of 23 Feb 2020

\bibitem{Abbott:2016nmj} B.~P.~Abbott {\it et al.} [LIGO Scientific and Virgo Collaborations],
  %``GW151226: Observation of Gravitational Waves from a 22-Solar-Mass Binary Black Hole Coalescence,''
  Phys.\ Rev.\ Lett.\  {\bf 116}, no. 24, 241103 (2016)
  doi:10.1103/PhysRevLett.116.241103
  [arXiv:1606.04855 [gr-qc]].
  %%CITATION = doi:10.1103/PhysRevLett.116.241103;%%
  %2121 citations counted in INSPIRE as of 10 Mar 2020

\bibitem{Abbott:2017oio} B.~P.~Abbott {\it et al.} [LIGO Scientific and Virgo Collaborations],
  %``GW170814: A Three-Detector Observation of Gravitational Waves from a Binary Black Hole Coalescence,''
  Phys.\ Rev.\ Lett.\  {\bf 119}, no. 14, 141101 (2017)
  doi:10.1103/PhysRevLett.119.141101
  [arXiv:1709.09660 [gr-qc]].
  %%CITATION = doi:10.1103/PhysRevLett.119.141101;%%
  %1155 citations counted in INSPIRE as of 10 Mar 2020

\bibitem{TheLIGOScientific:2017qsa} B.~P.~Abbott {\it et al.} [LIGO Scientific and Virgo Collaborations],
  %``GW170817: Observation of Gravitational Waves from a Binary Neutron Star Inspiral,''
  Phys.\ Rev.\ Lett.\  {\bf 119}, no. 16, 161101 (2017)
  doi:10.1103/PhysRevLett.119.161101
  [arXiv:1710.05832 [gr-qc]].
  %%CITATION = doi:10.1103/PhysRevLett.119.161101;%%
  %2993 citations counted in INSPIRE as of 10 Mar 2020

\bibitem{Monitor:2017mdv} B.~P.~Abbott {\it et al.} [LIGO Scientific and Virgo and Fermi-GBM and INTEGRAL Collaborations],
  %``Gravitational Waves and Gamma-rays from a Binary Neutron Star Merger: GW170817 and GRB 170817A,''
  Astrophys.\ J.\  {\bf 848}, no. 2, L13 (2017)
  doi:10.3847/2041-8213/aa920c
  [arXiv:1710.05834 [astro-ph.HE]].
  %%CITATION = doi:10.3847/2041-8213/aa920c;%%
  %1108 citations counted in INSPIRE as of 19 Feb 2020

\bibitem{Kelly:2014mwa} P.~L.~Kelly {\it et al.},
  %``Multiple Images of a Highly Magnified Supernova Formed by an Early-Type Cluster Galaxy Lens,''
  Science {\bf 347}, 1123 (2015)
  doi:10.1126/science.aaa3350
  [arXiv:1411.6009 [astro-ph.CO]].
  %%CITATION = doi:10.1126/science.aaa3350;%%
  %90 citations counted in INSPIRE as of 10 Mar 2020

\bibitem{Goobar:2016uuf} A.~Goobar {\it et al.},
  %``iPTF16geu: A multiply imaged, gravitationally lensed type Ia supernova,''
  Science {\bf 356}, 291 (2017)
  doi:10.1126/science.aal2729
  [arXiv:1611.00014 [astro-ph.CO]].
  %%CITATION = doi:10.1126/science.aal2729;%%
  %63 citations counted in INSPIRE as of 10 Mar 2020

\bibitem{Tanabashi:2018oca} M.~Tanabashi {\it et al.} [Particle Data Group],
  %``Review of Particle Physics,''
  Phys.\ Rev.\ D {\bf 98}, no. 3, 030001 (2018).
  %doi:10.1103/PhysRevD.98.030001
  %%CITATION = doi:10.1103/PhysRevD.98.030001;%%
  %2806 citations counted in INSPIRE as of 30 Oct 2019

\bibitem{Sakstein:2017xjx} J.~Sakstein and B.~Jain,
  %``Implications of the Neutron Star Merger GW170817 for Cosmological Scalar-Tensor Theories,''
  Phys.\ Rev.\ Lett.\  {\bf 119}, no. 25, 251303 (2017)
  %doi:10.1103/PhysRevLett.119.251303
  %[arxiv:1710.05893 [astro-ph.CO]].
  %%CITATION = doi:10.1103/PhysRevLett.119.251303;%%
  %254 citations counted in INSPIRE as of 12 Jun 2019

\bibitem{Baker:2017hug} T.~Baker, E.~Bellini, P.~G.~Ferreira, M.~Lagos, J.~Noller and I.~Sawicki,
  %``Strong constraints on cosmological gravity from GW170817 and GRB 170817A,''
  Phys.\ Rev.\ Lett.\  {\bf 119}, no. 25, 251301 (2017)
  %doi:10.1103/PhysRevLett.119.251301
  %[arxiv:1710.06394 [astro-ph.CO]].
  %%CITATION = doi:10.1103/PhysRevLett.119.251301;%%
  %200 citations counted in INSPIRE as of 17 Jan 2019

\bibitem{Fan:2016swi} X.~L.~Fan, K.~Liao, M.~Biesiada, A.~Piorkowska-Kurpas and Z.~H.~Zhu,
  %``Speed of Gravitational Waves from Strongly Lensed Gravitational Waves and Electromagnetic Signals,''
  Phys.\ Rev.\ Lett.\  {\bf 118}, no. 9, 091102 (2017)
  %doi:10.1103/physrevlett.118.091102, 10.1103/PhysRevLett.118.091102
  %[arxiv:1612.04095 [gr-qc]].
  %%CITATION = doi:10.1103/physrevlett.118.091102, 10.1103/PhysRevLett.118.091102;%%
  %24 citations counted in INSPIRE as of 09 May 2019

\bibitem{Wei:2017emo} J.~J.~Wei and X.~F.~Wu,
  %``Strongly Lensed Gravitational Waves and Electromagnetic Signals as Powerful Cosmic Rulers,''
  Mon.\ Not.\ Roy.\ Astron.\ Soc.\  {\bf 472}, no. 3, 2906 (2017)
  doi:10.1093/mnras/stx2210
  [arXiv:1707.04152 [astro-ph.CO]].
  %%CITATION = doi:10.1093/mnras/stx2210;%%
  %18 citations counted in INSPIRE as of 22 Feb 2020

\bibitem{Yang:2018bdf} T.~Yang, B.~Hu, R.~G.~Cai and B.~Wang,
  %``A new probe of gravity: strongly lensed gravitational wave multi-messenger approach,''
  Astrophys.\ J.\  {\bf 880}, 50 (2019)
  %arXiv:1810.00164 [astro-ph.CO].
  %%CITATION = ARXIV:1810.00164;%%

\bibitem{Jia:2019hih} J.~Jia and H.~Liu,
  %``Time delay of timelike particles in gravitational lensing of the Schwarzschild spacetime,''
  Phys.\ Rev.\ D {\bf 100}, no. 12, 124050 (2019)
  doi:10.1103/PhysRevD.100.124050
  [arXiv:1906.11833 [gr-qc]].
  %%CITATION = doi:10.1103/PhysRevD.100.124050;%%
  %2 citations counted in INSPIRE as of 09 Mar 2020

\bibitem{Bardeen:1968} J. Bardeen, Proceedings of GR5, eds. C. DeWitt and B. DeWitt, {\it Gordon and Breach} (1968), p. 174.

\bibitem{AyonBeato:2000zs} E.~Ayon-Beato and A.~Garcia,
  %``The Bardeen model as a nonlinear magnetic monopole,''
  Phys.\ Lett.\ B {\bf 493}, 149 (2000)
  doi:10.1016/S0370-2693(00)01125-4
  [gr-qc/0009077].
  %%CITATION = doi:10.1016/S0370-2693(00)01125-4;%%
  %246 citations counted in INSPIRE as of 13 Dec 2019

\bibitem{Janis:1968zz} A.~I.~Janis, E.~T.~Newman and J.~Winicour,
  %``Reality of the Schwarzschild Singularity,''
  Phys.\ Rev.\ Lett.\  {\bf 20}, 878 (1968).
  doi:10.1103/PhysRevLett.20.878
  %%CITATION = doi:10.1103/PhysRevLett.20.878;%%
  %263 citations counted in INSPIRE as of 14 Dec 2019

\bibitem{Breton:2002td} N.~Breton,
  %``Geodesic structure of the Born-Infeld black hole,''
  Class.\ Quant.\ Grav.\  {\bf 19}, 601 (2002).
  doi:10.1088/0264-9381/19/4/301
  %%CITATION = doi:10.1088/0264-9381/19/4/301;%%
  %41 citations counted in INSPIRE as of 05 Jan 2020

\bibitem{Eiroa:2005ag} E.~F.~Eiroa,
  %``Gravitational lensing by Einstein-Born-Infeld black holes,''
  Phys.\ Rev.\ D {\bf 73}, 043002 (2006)
  doi:10.1103/PhysRevD.73.043002
  [gr-qc/0511065].
  %%CITATION = doi:10.1103/PhysRevD.73.043002;%%

\bibitem{Weinberg:1972kfs} S.~Weinberg,
  {\it Gravitation and Cosmology: Principles and Applications of the General Theory of Relativity}, John Wiley \& Sons (1972)
  %%CITATION = INSPIRE-1410180;%%
  %184 citations counted in INSPIRE as of 04 Feb 2020

\bibitem{bk:inttable} I.S.Gradshteyn, and I.M.Ryzhik, \emph{Table of Integrals, Series, and Products}, 8th ed. Academic Press (2014), p. 152.

\bibitem{ourletter} Ke Huang and Junji Jia,
[arXiv:2003.08250v1 [gr-qc]].

\bibitem{Oguri:2006qp} M.~Oguri,
  %``Gravitational Lens Time Delays: A Statistical Assessment of Lens Model Dependences and Implications for the Global Hubble Constant,''
  Astrophys.\ J.\  {\bf 660}, 1 (2007)
  doi:10.1086/513093
  [astro-ph/0609694].
  %%CITATION = doi:10.1086/513093;%%
  %145 citations counted in INSPIRE as of 22 Feb 2020

\bibitem{Liao:2018ofi} K.~Liao, X.~Ding, M.~Biesiada, X.~L.~Fan and Z.~H.~Zhu,
  %``Anomalies in Time Delays of Lensed Gravitational Waves and Dark Matter Substructures,''
  Astrophys.\ J.\  {\bf 867}, no. 1, 69 (2018)
  doi:10.3847/1538-4357/aae30f
  [arXiv:1809.07079 [astro-ph.CO]].
  %%CITATION = doi:10.3847/1538-4357/aae30f;%%
  %5 citations counted in INSPIRE as of 22 Feb 2020

\bibitem{LIGOScientific:2018mvr} B.~P.~Abbott {\it et al.} [LIGO Scientific and Virgo Collaborations],
  %``GWTC-1: A Gravitational-Wave Transient Catalog of Compact Binary Mergers Observed by LIGO and Virgo during the First and Second Observing Runs,''
  Phys.\ Rev.\ X {\bf 9}, no. 3, 031040 (2019)
  doi:10.1103/PhysRevX.9.031040
  [arXiv:1811.12907 [astro-ph.HE]].
  %%CITATION = doi:10.1103/PhysRevX.9.031040;%%
  %722 citations counted in INSPIRE as of 23 Feb 2020

\bibitem{Sereno:2007rm} M.~Sereno,
  %``On the influence of the cosmological constant on gravitational lensing in small systems,''
  Phys.\ Rev.\ D {\bf 77}, 043004 (2008)
  doi:10.1103/PhysRevD.77.043004
  [arXiv:0711.1802 [astro-ph]].
  %%CITATION = doi:10.1103/PhysRevD.77.043004;%%
  %78 citations counted in INSPIRE as of 22 Feb 2020

\bibitem{Keeton:2005jd} C.~R.~Keeton and A.~O.~Petters,
  %``Formalism for testing theories of gravity using lensing by compact objects. I. Static, spherically symmetric case,''
  Phys.\ Rev.\ D {\bf 72}, 104006 (2005)
  doi:10.1103/PhysRevD.72.104006
  [gr-qc/0511019].
  %%CITATION = doi:10.1103/PhysRevD.72.104006;%%
  %108 citations counted in INSPIRE as of 24 Feb 2020

\bibitem{Glicenstein:2017lrm} J.~F.~Glicenstein,
  %``Gravitational lensing time delays with massive photons,''
  Astrophys.\ J.\  {\bf 850}, no. 1, 102 (2017)
  doi:10.3847/1538-4357/aa9439
  [arXiv:1710.11587 [astro-ph.HE]].
  %%CITATION = doi:10.3847/1538-4357/aa9439;%%
  %3 citations counted in INSPIRE as of 24 Feb 2020

\bibitem{Akiyama:2019cqa}
  K.~Akiyama {\it et al.} [Event Horizon Telescope Collaboration],
  %``First M87 Event Horizon Telescope Results. I. The Shadow of the Supermassive Black Hole,''
  Astrophys.\ J.\  {\bf 875}, no. 1, L1 (2019)
  doi:10.3847/2041-8213/ab0ec7
  [arXiv:1906.11238 [astro-ph.GA]].
  %%CITATION = doi:10.3847/2041-8213/ab0ec7;%%
  %379 citations counted in INSPIRE as of 16 Mar 2020

\end{thebibliography}
\end{document}